\documentclass[twocolumn,prl,aps,psfig,showpacs,preprintnumbers,superscriptaddress]{revtex4}
\usepackage{graphicx}
\usepackage{epstopdf}
\usepackage{float}
\usepackage{color}
\usepackage{amsmath}

\begin{document}

\title{Competing Magnetic Fluctuations in Iron Pnictide Superconductors: Role of Ferromagnetic Spin Correlations Revealed by NMR }

\author{P.~Wiecki}
\affiliation{The Ames Laboratory and Department of Physics and Astronomy, Iowa State University, Ames, Iowa 50011, USA}
\author{B.~Roy}
\affiliation{The Ames Laboratory and Department of Physics and Astronomy, Iowa State University, Ames, Iowa 50011, USA}
\author{D.~C.~Johnston}
\affiliation{The Ames Laboratory and Department of Physics and Astronomy, Iowa State University, Ames, Iowa 50011, USA}
\author{S.~L.~Bud'ko}
\affiliation{The Ames Laboratory and Department of Physics and Astronomy, Iowa State University, Ames, Iowa 50011, USA}
\author{P.~C.~Canfield}
\affiliation{The Ames Laboratory and Department of Physics and Astronomy, Iowa State University, Ames, Iowa 50011, USA}
\author{Y.~Furukawa}
\affiliation{The Ames Laboratory and Department of Physics and Astronomy, Iowa State University, Ames, Iowa 50011, USA}

\date{\today}

\begin{abstract}
     In the iron pnictide superconductors, theoretical calculations have consistently shown enhancements of the static magnetic susceptibility at both the stripe-type antiferromagnetic (AFM) and in-plane ferromagnetic (FM) wavevectors. 
      However, the possible existence of FM fluctuations has not yet been examined from a microscopic point of view. 
      Here, using  $^{75}$As NMR data, we provide clear evidence for the existence of FM spin correlations in both the hole- and electron-doped BaFe$_2$As$_2$ families of iron-pnictide superconductors. 
     These FM fluctuations appear to compete with superconductivity and are thus a crucial ingredient to understanding the variability of $T_{\rm c}$ and the shape of the superconducting dome in these and other iron-pnictide families.
\end{abstract}

\maketitle
   The role of magnetic fluctuations in iron pnictide superconductors (SCs) has been extensively studied since their discovery.
    As the parent materials have antiferromagnetic (AFM) ground states, attention has been understandably focused on stripe-type AFM fluctuations,
   which are widely believed to give rise to the Cooper pairing in these systems. 
    In the standard picture, carrier doping or pressure application results in suppression of the AFM order and the emergence of a SC state, with $T_{\rm c}$ ranging from a few K to 56 K \cite{Johnston2010}.
   However, as of yet, there is no accepted theory for $T_{\rm c}$ in these materials with which to explain the large variability in maximum $T_{\rm c}$ between different iron arsenide families and the different shapes of the SC dome with electron and hole doping.

   Recent nuclear magnetic resonance (NMR) measurements on non-SC, paramagnetic (PM) SrCo$_2$As$_2$, the $x=1$ member of the electron-doped  Sr(Fe$_{1-x}$Co$_x$)$_2$As$_2$ family, revealed strong ferromagnetic (FM) spin fluctuations in the Co layer coexisting with stripe-type AFM fluctuations \cite{Wiecki2015,Jayasekara2013}. 
     Since stripe-type AFM fluctuations are a key ingredient to SC in the iron pnictides, this result suggested that FM fluctuations might compete with the stripe-type AFM fluctuations, suppressing SC in SrCo$_2$As$_2$. 
   FM correlations were also observed in isostructural BaCo$_2$As$_2$ \cite{Wiecki2015,Ahilan2014}.
   Similarly, CaCo$_{1.86}$As$_2$ has an A-type AFM ground state with in-plane FM order \cite{Quirinale2013}.
   These results also raise the question of whether similar FM correlations exist generally in the SC $A$(Fe$_{1-x}$Co$_x$)$_2$As$_2$ compounds, not just at the $x=1$ edges of their phase diagrams.

     According to density functional theory calculations \cite{Singh2008,Mazin2008,Dong2008,Yaresko2009,Neupane2011}, the generalized 
static magnetic susceptibility $\chi(\mathbf{q})$ is enhanced at both the FM and stripe-type AFM wavevectors in all the iron-based SCs and parent compounds. 
     Experimentally, the uniform $\chi(\mathbf{q}=0)$ of the parent compounds is enhanced by a factor
of order five over band structure values, which is consistent with FM correlations \cite{Johnston2010}. 
     Nevertheless, FM fluctuations have not been investigated microscopically, perhaps because low-energy FM fluctuations are difficult to observe via inelastic neutron scattering (INS). 
     The peak in the inelastic structure factor at $\mathbf{q}=0$ coincides with the elastic Bragg diffraction peaks, and the energy scale of thermal neutrons is relatively high. 
     The study of low-energy FM fluctuations therefore requires cold, polarized neutrons.  
     NMR, in contrast, is a microscopic probe uniquely sensitive to low-energy FM fluctuations via the modified Korringa ratio.

     In this Letter, using $^{75}$As NMR measurements, we present clear evidence for FM fluctuations in the tetragonal, PM phase of both the hole- and electron-doped BaFe$_2$As$_2$ families of iron pnictide SCs. 
    Furthermore, we suggest that these FM fluctuations compete with SC, and that this competition between FM and AFM fluctuations may be a key ingredient to a theory of $T_{\rm c}$ in the iron pnictides. 

    For this study, we chose $x=4.7$ $\%$ ($T_{\rm N}\sim 50$  K and $T_{\rm c}\sim 15$ K) and $x=5.4$  $\%$ ($T_{\rm N}\sim 35$ K and $T_{\rm c}\sim 20$ K) in single-crystalline  Ba(Fe$_{1-x}$Co$_x$)$_2$As$_2$ as representative superconducting samples in which to look for FM correlations.
    We also used our existing data on BaCo$_2$As$_2$, reported elsewhere \cite{Anand2014,Wiecki2015} and other data from the literature.
   The $^{75}$As NMR shift and spin-lattice relaxation rates $1/T_1$ were measured under magnetic fields parallel to the $c$ axis ($H \| c$) and  to the $ab$ plane ($H \| ab$).

\begin{figure*}[t]
\centering
\includegraphics[width=18cm]{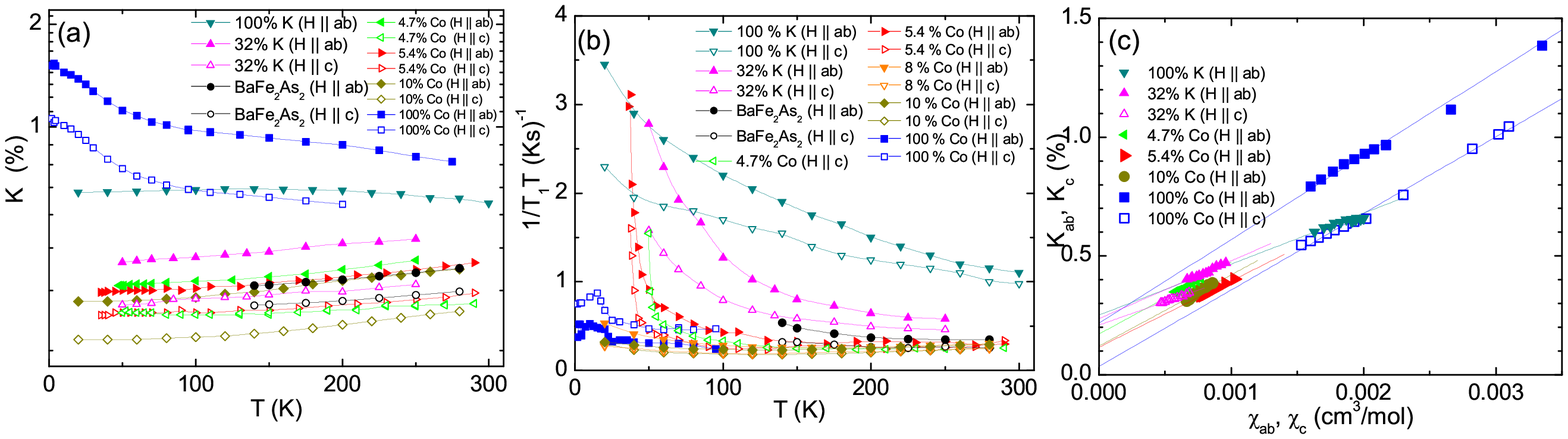}
\caption{(Color online) (a) $T$ dependence of the NMR shift $K$ for a variety of indicated samples. 
(b) $T$ dependence of NMR spin-lattice relaxation rate $1/T_1T$ for the same samples. 
Here, and throughout, filled (open) symbols are used for $H\|ab$ ($H\|c$). 
(c) $K_i(T)$ vs. $\chi_i(T)$ ($i=ab,c$)  for a variety of samples.
Data are from Refs. \cite{Ning2010,Ning2009b, Kitagawa2008,Ning2008,Hirano2012,Ni2008,Sefat2008,Liu2014}.
For KFe$_2$As$_2$, $K_c$ is nearly $T$ independent making a $K$ vs. $\chi$ analysis impossible. 
From the relative values of $K_{ab}$ and $K_c$, we estimate $K_{0,c}\sim0.21\%$. 
For  BaFe$_2$As$_2$, $K_{0,ab}=0.14\%$ and $K_{0,c}=0.21\%$ are from Ref. \onlinecite{Kitagawa2008}.  For the 8$\%$ Co doped sample, $K_{0,c}=0.22\%$ from  Ref. \onlinecite{Ning2010}.
}
\label{fig:data}
\end{figure*}

   Figures~\ref{fig:data}(a) and \ref{fig:data}(b) show the existing NMR data ($T$ dependence of NMR shift $K$ and $1/T_1T$, respectively) for both the electron-doped Ba(Fe$_{1-x}$Co$_x$)$_2$As$_2$ and hole-doped Ba$_{1-x}$K$_x$Fe$_2$As$_2$ families of iron-pnictide SCs.
     The NMR shift consists of a $T$-independent orbital shift $K_0$ and a $T$-dependent spin shift $K_{\text{spin}}(T)$ due to the uniform magnetic spin susceptibility $\chi(\mathbf{q}=0)$.
     The NMR shift can therefore be expressed as $K(T)=K_0+K_{\text{spin}}(T)=K_0+A_{\text{hf}}\chi_{\text{spin}}/N_{\rm A}$, where $N_{\rm A}$ is Avogadro's number, and 
$A_{\text{hf}}$ is the hyperfine coupling, usually expressed in units of kOe$/\mu_{\rm B}$. 
    In order to extract $K_{\text{spin}}(T)$, we plot $K(T)$ against the corresponding bulk static uniform magnetic susceptibility $\chi(T)$ with $T$ as shown in Fig. \ref{fig:data}(c).
    From the $y$-intercept of the linear fit curve we can estimate the orbital shift $K_0$, and extract $K_{\text{spin}}(T)$ needed for the following Korringa ratio analysis.
    
      To extract the character of spin fluctuations in the iron pnictides from $^{75}$As NMR data, we employ a modified Korringa ratio analysis. 
      Within a Fermi liquid picture, both $1/T_1T$ and $K_{\text{spin}}$ are determined primarily by the density of states at the Fermi energy ${\cal D}(E_{\rm F})$, leading to the Korringa relation  $T_1TK_{\text{spin}}^2$  = $(\hbar/4\pi k_{\rm B})\left(\gamma_{\rm e}/\gamma_{\rm N}\right)^2 \equiv S$.
    For the $^{75}$As nucleus ($\gamma_{\rm N}/2\pi=7.2919$ MHz/T),  $S=8.97\times 10^{-6}$ Ks. 
    Deviations from $T_1TK_{\text{spin}}^2=S$, which are conveniently expressed via the Korringa ratio $\alpha\equiv S/(T_1TK_{\text{spin}}^2)$, can reveal information about electron correlations in the material \cite{Moriya1963,Narath1968}.
    For uncorrelated electrons, we have $\alpha\sim1$. 
     However, enhancement of $\chi(\mathbf{q}\neq 0)$ increases $1/T_1T$ but has little or no effect on $K_{\text{spin}}$, which probes only the uniform $\chi(\mathbf{q} = 0)$.  
    Thus $\alpha >1$ for AFM correlations. In contrast, $\alpha <1$ for FM correlations.
    The Korringa ratio $\alpha$, then, reveals whether the magnetic correlations in the material have predominantly FM or AFM character.

  To perform the Korringa ratio analysis, one needs to take the anisotropy of $K_\text{spin}$ and $1/T_1T$ into consideration.
   The $1/T_1$ probes hyperfine field fluctuations at the NMR Larmor frequency, $\omega_{\rm N}$, perpendicular to the external magnetic field according to $(1/T_1)_{H||i}=\gamma_{\rm N}^2\left[|H^{\rm hf}_j(\omega_{\rm N})|^2+|H^{\rm hf}_k(\omega_{\rm N})|^2\right]$,
where $(i,j,k)$ are mutually orthogonal directions and $|H^{\rm hf}_j(\omega)|^2$ represents the power spectral density of the $j$-th component of the hyperfine magnetic field at the nuclear site. 
    Thus, defining $H^{\rm hf}_{ab}\equiv H^{\rm hf}_{a}=H^{\rm hf}_{b}$, which is appropriate for the tetragonal PM state, we have $(1/T_1)_{H||c}=2\gamma_{\rm N}^2|H^{\rm hf}_{ab}(\omega_{\rm N})|^2\equiv 1/T_{1,\perp}$. 
    The Korringa parameter $\alpha_{\bot}\equiv S/T_{1,\bot}TK_{\text{spin},ab}^2$ will then characterize fluctuations in the $ab$-plane component of the hyperfine field. 
    By analogy, we should pair $K_{\text{spin},c}$ with $2\gamma_N^2|H^{\rm hf}_{c}(\omega_{\rm N})|^2\equiv 1/T_{1,\|}$, so that the Korringa parameter
    $\alpha_{\|}=S/T_{1,\|}TK_{\text{spin},c}^2$ characterizes fluctuations in the $c$-axis component of the hyperfine field. 
    Since $(1/T_1)_{H||ab}=\gamma_N^2\left[|H^{\rm hf}_{ab}(\omega_{\rm N})|^2+|H^{\rm hf}_c(\omega_{\rm N})|^2\right]$, we estimate the quantity $1/T_{1,\|}T$ from $1/T_{1,\|}T=2(1/T_1T)_{H||ab}-(1/T_1T)_{H||c}$.

\begin{figure}[t]
\centering
\includegraphics[width=\columnwidth]{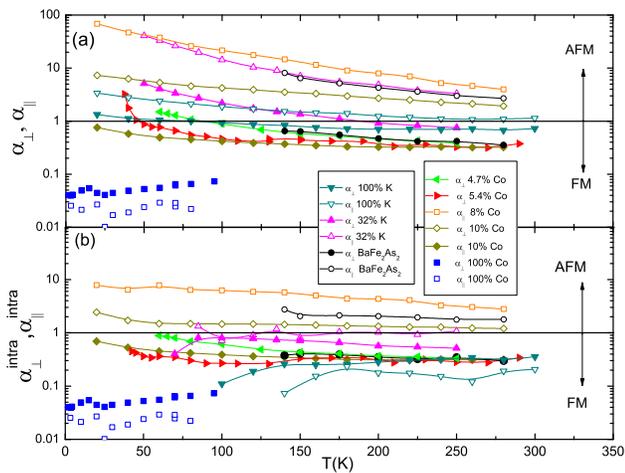}
\caption{(Color online) (a) Korringa ratios $\alpha_\perp$ (filled symbols) and $\alpha_\|$ (open symbols) 
as a function of $T$ in a variety of Ba(Fe$_{1-x}$Co$_x$)$_2$As$_2$ and Ba$_{1-x}$K$_x$Fe$_2$As$_2$ samples. 
    (b) Same for intraband Korringa ratios $\alpha_\|^\text{intra}$ and $\alpha_\perp^\text{intra}$, obtained by subtracting the interband (Curie-Weiss) contributions.  }
\label{fig:alpha}
\end{figure}

The $T$ dependences of the Korringa ratios $\alpha_{\bot}=S/T_{1,\bot}TK_{\text{spin},ab}^2$ and $\alpha_{\|}=S/T_{1,\|}TK_{\text{spin},c}^2$ are shown in Fig. \ref{fig:alpha}(a).
   In BaCo$_2$As$_2$, both $\alpha_\perp$ and $\alpha_\|$ are nearly independent of $T$  and much less than 1, consistent with FM correlations.
   For the remaining samples,  $\alpha_\|$ is generally greater than 1 indicating AFM correlations throughout the $T$ range. 
   In addition, both $\alpha_\perp$ and $\alpha_\|$ increase as $T$ is lowered, showing the growth of AFM spin fluctuations at low $T$.
   In contrast,  we find that $\alpha_\perp\sim0.3<1$ for the parent and Co-doped samples in the high-$T$ PM phase.
   The hole-doped Ba$_{1-x}$K$_x$Fe$_2$As$_2$  also display $\alpha_\perp\leq1$ in the PM phase, suggesting FM correlations, although less strong than in the Co-doped samples.

    Careful consideration is required to interpret the value of the Korringa ratio.
    In comparing the $\alpha$ value to the crossover $\alpha_0=1$ between dominant FM and AFM fluctuations, one is assuming a simple model in which the nuclear 
relaxation is due to the local ${\cal{D}}(E_F)$ at the As sites through on-site hyperfine
interactions, where As-$4p$ bands hybridize with Fe-$3d$ bands \cite{Kitagawa2008}. 
   If, on the other hand, the nuclear relaxation is induced only by the localized Fe spins through isotropic transferred hyperfine interactions, the value of $\alpha$ would instead be compared to the crossover $\alpha_0=1/4$, assuming no contributions to 1/$T_1$ from AFM correlations due to form factor effects \cite{Millis1990,Jeglic2010,Lang2008}.
  In the highly overdoped $x=26\%$ Ba(Fe$_{1-x}$Co$_x$)$_2$As$_2$, however, AFM fluctuations are known to be absent from INS measurements \cite{Matan2010}. 
   Accordingly, Refs. \onlinecite{Johnston2010} and \onlinecite{Ning2010} find $\alpha\sim1.2$, suggesting weak correlation. 
    If the crossover were $\alpha_0=1/4$, this value of 1.2 for the Korringa ratio must be associated with dominant AFM fluctuations, in conflict with observations. 
    These results suggest that the factor of 4 change to $\alpha_0 = 1/4$ proposed by Ref. \cite{Jeglic2010} for iron pnictides is too large.
    In fact, the FM correlations have been also pointed out in (La$_{0.87}$Ca$_{0.13}$)FePO with $\alpha$ = 0.37 by $^{31}$P NMR \cite{Nakai2008}. 
     In addition, in the case of Na$_x$CoO$_2$ for $x$ $>$ 0.65, FM correlations are known to be present \cite{Lang2008} and the measured Korringa ratio takes the value $\alpha$ $\sim$ 0.3 \cite{Alloul2008}. 
     It is also noted that the Wilson ratio for BaFe$_2$As$_2$ is mildly enhanced ($R_{\rm W}$ $\sim$  3) \cite{Sefat2009}, consistent with FM correlations.  
   Thus we conclude that value we observe, $\alpha_\perp\sim0.3$, can be reasonably attributed to FM fluctuations.

\begin{figure}[b]
\centering
\centering
\includegraphics[width=8.7cm]{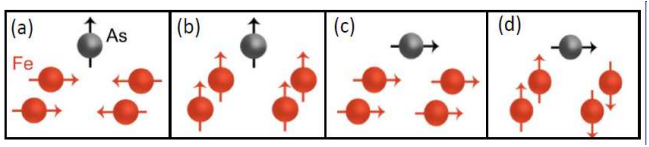}
\caption{(Color online) (a),(b): Competing sources of hyperfine field fluctuations along the $c$ axis. 
(c),(d): Competing sources of hyperfine field fluctuations in the $ab$ plane. 
 Competition between (a) and (b) [(c) and (d)] determines the value of $\alpha_\|$ ($\alpha_\perp$).  
}
\label{fig:hyperfine}
\end{figure}

    To discuss the magnetic correlations based on the values of $\alpha_\perp$ and $\alpha_\|$ in the iron pnictides in more detail, it is helpful to consider the hyperfine field at the $^{75}$As site, which is determined by the spin moments on the Fe sites through the hyperfine coupling tensor \cite{Hirano2012}.
    In this case, there are two sources of hyperfine field pointing along the $c$ axis \cite{Kitagawa2008}: stripe-type AFM fluctuations at $\mathbf{q}=(\pi,0)/(0,\pi)$ with the spins pointing within the $ab$ plane [as illustrated in Fig. \ref{fig:hyperfine}(a)] or FM fluctuations at $\mathbf{q}=0$ with the spins pointing along the $c$ axis [Fig.~\ref{fig:hyperfine}(b)].
   Similarly, hyperfine field fluctuations in the $ab$ plane can result from FM fluctuations at $\mathbf{q}=0$ with the spins pointing within the $ab$ plane [Fig. \ref{fig:hyperfine}(c)], or from AFM fluctuations at $\mathbf{q}=(\pi,0)/(0,\pi)$ with the spins pointing along the $c$ axis [Fig. \ref{fig:hyperfine}(d)].
   Thus, the value of $\alpha_\|$ reflects the competition between (a)- and (b)-type correlations: type (a) AFM correlations will increase $\alpha_\|$ above 1, while type (b) FM correlations will lower $\alpha_\|$ below 1. 
  Similarly,  $\alpha_\perp$ reflects the competition between (c)- and (d)-type correlations:
type (d) AFM correlations will increase $\alpha_\perp$, while type (c) FM correlations will lower $\alpha_\perp$.
   In what follows, we will refer to the correlations depicted in Fig. \ref{fig:hyperfine}(a) as ``type (a)'' correlations (similarly for the others).
   Since $\alpha_\|$ reflects the character of the $c$-axis component hyperfine field fluctuations, the AFM value of $\alpha_\|$ in Fig.~\ref{fig:alpha} can be attributed to type (a) correlations, i.e., stripe-type AFM correlations with the Fe spins in-plane. 
   These must dominate type (b) fluctuations in order to have an AFM value of $\alpha_\|$.
   Similarly, since $\alpha_\perp$ reflects the character of the $ab$-plane component of hyperfine field fluctuations, the  FM value of $\alpha_\perp$ in the high-$T$ region can be attributed to type (c) in-plane FM fluctuations.
   On the other hand, the increase of $\alpha_\perp$ as the temperature is lowered reflects the increasing dominance of type (d) stripe-type AFM correlations with a $c$-axis component to the spin. 
    This clearly indicates the simultaneous coexistence of FM and AFM fluctuations.
   Furthermore, the dominance of type (a) and (c) spin fluctuations in the high-$T$ region suggests that both the AFM and FM fluctuations are highly anisotropic in the iron pnictides, favoring the $ab$ plane.

\begin{figure}[t]
\centering
\includegraphics[width=\columnwidth]{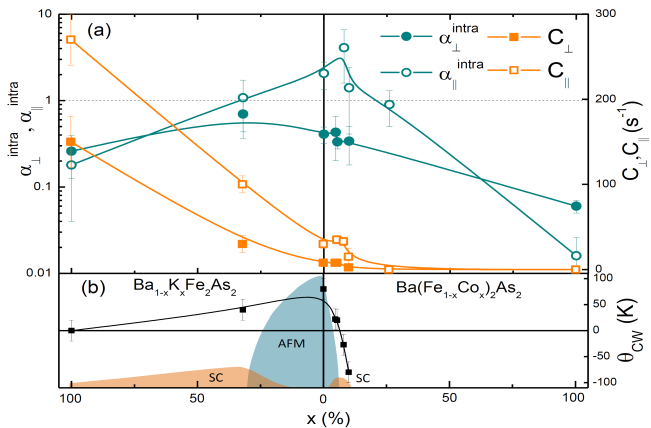}
\caption{(Color online) Potential relevance of FM spin fluctuations to iron pnictide phase diagram: 
    (a) Doping dependence of the nearly $T$-independent values of $\alpha_\|^\text{intra}$ and $\alpha_\perp^\text{intra}$, which parameterize the strength of FM fluctuations \cite{x=26}. 
We also show the doping dependence of the Curie-Weiss parameters $C_\perp$ and $C_\|$, which parameterize the strength of 2D AFM fluctuations. 
    (b) The doping dependence of the Curie-Weiss temperature $\Theta_{\rm CW}$. Solid lines are guides to the eye.}
\label{fig:phase}
\end{figure}

   Finally it is interesting to isolate the FM fluctuations and extract their $T$ dependence.
   We adopt the simple phenomenological model of Refs.~\cite{Ning2010,Ahilan2009,Nakai2013} to decompose 1/$T_1T$ into inter- and intraband components according to $1/T_1T=(1/T_1T)_\text{inter}+(1/T_1T)_\text{intra}$.
  The $T$ dependence of the interband term is assumed to follow the Curie-Weiss form appropriate for 2D AFM fluctuations: $(1/T_1T)_\text{inter}=C/(T-\Theta_{\rm CW})$.
   For the Co-doped samples, we use $(1/T_1T)_\text{intra}=\alpha+\beta\text{exp}(-\Delta/k_{\rm B}T)$, while for the K-doped samples we simply use $(1/T_1T)_\text{intra}=$ const, as in Ref.~\onlinecite{Hirano2012}. 
    The Curie-Weiss parameter $C$ measures the strength of AFM fluctuations, and  $\Theta_{\rm CW}$ corresponds to the distance in $T$ from the AFM instability point.
    Here, we decompose the quantities $1/T_{1,\|}T$ and $1/T_{1,\perp}T$  into their inter- and intraband components. 
Our results for the CW parameters $C_\perp$, $C_\|$  and $\Theta_{\rm CW}$, shown in Fig. \ref{fig:phase}, are consistent with the results of Refs.~\onlinecite{Hirano2012} and \onlinecite{Ning2010}. 
    Similar carrier doping dependence of $\Theta_{\rm CW}$ is reported in P-doped BaFe$_2$As$_2$ \cite{Nakai2010_2} and in  LaFeAsO$_{1-x}$F$_x$ \cite{Oka2012}.
    We use the intraband components to calculate the Korringa ratios $\alpha_\|^\text{intra}$ and  $\alpha_\perp^\text{intra}$. 
   The results are shown in Fig. \ref{fig:alpha}(b).
    Both $\alpha_\|^\text{intra}$ and $\alpha_\perp^\text{intra}$ remain roughly constant through the $T$ range. 
   The deviations at low $T$ are due to imperfect subtraction of the interband part, arising from our simplistic Curie-Weiss fitting. 
    We notice that $\alpha_\|^\text{intra}$ for several compounds are greater than 1, suggesting AFM correlations in the intraband component.  
   On the other hand, the value of $\alpha_\perp^\text{intra}$ is consistent with FM fluctuations, as discussed above, for all samples.

    What then is the role of these FM fluctuations in the iron pnictide superconductors?  
    In Fig.~\ref{fig:phase}, we summarize our results across the combined hole- and electron-doped phase diagram of BaFe$_2$As$_2$.
     First of all, $C_\|$ is always greater than $C_\perp$ in the entire phase diagram, indicating that type (a) spin fluctuations are stronger than type (d) spin fluctuations.
    On the electron-doped side, AFM  spin fluctuations die out beyond the SC dome at $x\sim15\%$ \cite{Matan2010}. 
    In contrast, the AFM spin fluctuations become very strong on the hole-doped side relative to the electron-doped side.
    The doping dependences of $C_\perp$ and $C_\|$ are reminiscent of the doping dependence of the mass enhancement \cite{Medici2014}.  
    For a measure of the strength of the FM fluctuations, we plot in Fig. \ref{fig:phase}(a) the average values of  $\alpha_\|^\text{intra}$ and $\alpha_\perp^\text{intra}$ above 150 K except for BaCo$_2$As$_2$ where we average over all data.
    We find that $\alpha_\perp^\text{intra}$ has a FM value throughout the phase diagram, consistent with in-plane FM [type (c)] spin fluctuations. 
     In contrast, $\alpha_\|^\text{intra}$ shows an AFM value at 8$\%$ Co doping, but exhibits a dramatic decrease towards FM values when hole doped or electron doped beyond 8$\%$. 
The FM fluctuations are thus strongest at the maximally-doped edges of the phase diagram.
   The disappearance of AFM spin fluctuations beyond 15$\%$ Co doping coincides with the appearance of FM fluctuations, suggesting a competition between FM and AFM fluctuations. 
   On the hole-doped side, AFM correlations clearly increase in strength. 
   Paradoxically, this increase in strength of AFM correlations is accompanied by a decrease of $T_{\rm c}$, as noted in Ref. \onlinecite{Hardy2013}.
   Our analysis offers a possible explanation. FM correlations also increase in strength on the hole-doped side, as seen from the rapidly decreasing values of 
   $\alpha_\|^\text{intra}$ and $\alpha_\perp^\text{intra}$ and the increasing value of the NMR shift [Fig. \ref{fig:data}(a)] with increasing hole doping. 
   We suggest that the growth of competing FM correlations results in the reduction of $T_{\rm c}$ despite the increase in AFM correlation strength.
   In KFe$_2$As$_2$, then, FM and AFM correlations coexist with neither dominating the other, leading to  the Korringa parameters $\alpha_\perp\sim1$ and $\alpha_\|\sim1$ that we observe in Fig. \ref{fig:alpha}(a).
    Finally, it is noted that structural parameters have been pointed out to play an important role for controlling the ground state of iron pnictides \cite{Kimber2009}. 
    Although we discussed our NMR data based on the well-known phase diagram where the tuning parameter is carrier doping, the observed trends should not be attributed to carrier concentration alone.

    In conclusion, using an anisotropic modified Korringa ratio analysis on $^{75}$As NMR data, we have provided clear evidence for the existence of FM spin correlations in both hole- and  electron-doped BaFe$_2$As$_2$. 
   The FM fluctuations are strongest in the maximally-doped BaCo$_2$As$_2$ and KFe$_2$As$_2$, but are still present in the
BaFe$_2$As$_2$ parent compound, consistent with its enhanced $\chi$ \cite{Johnston2010}. 
    While we consider here only the Ba122 system, similar results are found for other iron-pnictide based superconductors. 
    In particular, FM values of $\alpha$ were also observed in the PM phase of LaO$_{0.9}$F$_{0.1}$FeAs ($\alpha=0.55<1$) \cite{Grafe2008}, K$_{0.8}$Fe$_2$Se$_2$ ($\alpha=0.45<1$) \cite{Kotegawa2011} and Ca(Fe$_{1-x}$Co$_x$)$_2$As$_2$ \cite{Cui}.
   These FM fluctuations appear to compete with superconductivity and are thus a crucial ingredient to understand the variability of  $T_{\rm c}$ and the shape of the SC dome. 
    Our results indicate that theoretical microscopic models should include FM correlations to capture the phenomenology of the iron pnictides.   
    Polarized INS experiments examining magnetic response at the FM wavevector will be needed to further understand the interplay between FM and AFM spin correlations in the iron pnictides.

The authors would like to acknowledge N. Ni for working on growth and basic characterization of the Co-substituted samples.
The research was supported by the U.S. Department of Energy, Office of Basic Energy Sciences, Division of Materials Sciences and Engineering. Ames Laboratory is operated for the U.S. Department of Energy by Iowa State University under Contract No.~DE-AC02-07CH11358.


\begin{thebibliography}{99}
\bibitem{x=26}    
 For the point at $x=26\%$, we used the data from Ref. \onlinecite{Ning2010} with $K_{0,c}=0.15\%$, and approximated $1/T_{1,\|}T\sim (1/T_1T)_{H||ab}$ as $(1/T_1T)_{H||c}$ data were not available.
\end{thebibliography}
\end{document}